\documentclass[useAMS,usenatbib,usegraphicx]{mn2e}

\title[Gamma-ray burst in AD 774/5]{A Galactic short gamma-ray burst as cause for the $^{14}$C peak in AD 774/5}

\author[V.V. Hambaryan \& R. Neuh\"auser]{V.V. Hambaryan$^{1}$\thanks{E-mail:
vvh@astro.uni-jena.de},
and R. Neuh\"auser$^{1}$ \\
$^{1}$Astrophysikalisches Institut, Universit\"at Jena, Schillerg\"asschen 2-3, 07745 Jena, Germany}

\begin{document}

\date{Accepted . Received ; in original form 2012}

\pagerange{\pageref{firstpage}--\pageref{lastpage}} \pubyear{2012}

\maketitle

\label{firstpage}

\begin{abstract}
In the last 3000 yr, one significant and rapid increase in the concentration of $^{14}$C in tree rings was observed;
it corresponds to a $\gamma$-ray energy input of $7 \cdot 10^{24}$ erg at Earth within up to one year in AD 774/5 \citep{m5}.
A normal supernova and a solar or stellar flare are unlikely as cause \citep{m5}, so that the source remained unknown. 
Here, we show that a short gamma-ray burst (GRB) in our Galaxy is consistent with all observables: 
Such an event is sufficiently short and provides the necessary energy in the relevant spectral range of $\gamma$-rays. 
Its spectral hardness is consistent with the differential production rates of $^{14}$C and $^{10}$Be as observed. 
The absence of reports about a historic sighting of a supernova in AD 774/5 or a present-day supernova remnant 
are also consistent with a short GRB. We estimate the distance towards this short GRB to be $\sim 1$ to $4$ kpc -
sufficiently far away, so that no extinction event on Earth was triggered. 
This is the first evidence for a short GRB in our Galaxy. 
\end{abstract}


\section{Introduction: The AD 774/5 event}

A significant increase in the $^{14}$C to $^{12}$C isotope ratio was detected in Japanese trees 
in AD 774/5 and a subsequent decrease for $\sim 10$ yr \citep[henceforth M1]{m5}.
It is consistent with an increase in $^{14}$C in American and European trees with $5$ to $10$ yr time resolution \citep{s3}.
If deposited within one year or less - best consistent with an atmospheric deposition model - the increase is 
10 times larger than the average production due to Galactic cosmic rays and 20 times larger than expected from 
the $2 \times 11$ yr solar cycle (M12). This requires a $\gamma$-ray energy input of $7 \cdot 10^{24}$  erg at Earth (M12). 
Also, a $30~\%$ increase in $^{10}$Be around AD 775 was observed in Antarctica, but with lower time resolution \citep{h1}.
Solar or stellar flare were found to be unlikely because of the insufficient energetics and spectrum of 
such flares (M12). A normal supernova (SN) was also found to be unlikely from the lack of 
any historical sighting or a SN remnant (SNR) (M12). 

\section{Supernova or magnetar flare~?}

A strongly absorbed SN was not considered quantitatively, yet: Absorption in the line-of-sight would not affect $\gamma$-rays, 
but would decrease the observable optical flux of a SN. 
Of the total energy output of a SN, E(event) = $10^{51}$ erg, a fraction $g = 0.01$ goes into $\gamma$-rays \citep{r3}. 
The ratio between the $\gamma$-ray energy emitted by a SN event 
spread homogeneously into the total area of a spherical shell around the SN ($4\cdot \pi \cdot d^{2}$ with distance $d$ 
from the event to Earth) and the $\gamma$-ray energy E(obs) observed at Earth is equal to the ratio 
between the surface area of that sphere and the Earth solid angle $\pi \cdot R^{2}$ (with Earth radius $R$):

\begin{equation}
\frac{E(event) \cdot g}{E(obs)} = \frac{4 \cdot \pi \cdot d^{2}}{\pi \cdot R^{2}} 
\end{equation}

Therefore, a normal SN (with $g = 0.01$), 
of which a $\gamma$-ray flux of E(obs) = $7 \cdot 10^{24}$ erg was observed at Earth, 
would have a distance $d \simeq 124$ pc, independent of absorption. 
If the AD 774/5 event were one of the rare ($1~\%$) over-luminous SNe, up to four times brighter 
than normal SNe \citep{r3}, then the expected distance is $d \simeq 260$ pc. 
From the peak absolute magnitude $M$ \citep{r3}, we can estimate the unabsorbed apparent peak magnitude:  \\
m=$-14.0 \pm 0.5$ mag, 124 pc, M=$-19.5 \pm 0.5$ mag, SN Ia, \\
m=$-12.5 \pm 1.0$ mag, 124 pc, M=$-18.0 \pm 1.0$ mag, SN II, \\
m=$-13.2 \pm 0.3$ mag, 260 pc, M=$-20.3 \pm 0.3$ mag, SN Ibc.  \\
One would need absorption of at least $A_{\rm V} \simeq 13$ mag 
to disable a historical sighting by naked eye \citep[limit $m \simeq 2$ mag for discovery of a new object,][]{s2}.
Such a strong absorption within $\sim 124$ or $260$ pc is not possible, 
except in small areas towards dark clouds \citep{r1}:
The closest dark clouds with $A_{\rm V} \ge 13$ mag are
Lynds 183 at $\sim 110$ pc with up to A$_{\rm V} = 150$ mag \citep{p1}
and $\rho$ Oph at $\sim 119$ pc with up to A$_{\rm V} = 65$ mag \citep{l2,s1}.
Absorption of $A_{\rm V} \ge 13$ mag is limited to $56$ deg$^{2}$ on the sky \citep[and K. Dobashi priv. comm.]{d2}
and less for distances within $124$ to $260$ pc \citep{r1}. The probability of an event within $56$ deg$^{2}$ 
of the whole sky is $0.0013$. Even then, a large, young, and bright SNR would be
detectable by X-ray pointings, but can be excluded \citep[Chandra SNR catalog\footnote{hea-www.harvard.edu/ChandraSNR/snrcat$\_$gal.html}]{g2}.

Given the measurement precision achieved in $^{14}$C for the $7.2~\sigma$ peak in AD 774/5 (M12), 
potential $^{14}$C from SNe can be detected up to $\sim 200$ pc with $3~\sigma$. Indeed, there are no SNe, pulsars, 
nor SNR known within a few hundred pc with age of some 300 to 2000 yr \citep[footnote 1; McGill SGR/AXP catalog\footnote{www.physics.mcgill.ca/~pulsar/magnetar/main.html}]{s2,g2,m1}
There are eleven events with evidence (historic observation, detected SNR and/or known pulsar) 
for a SN within 2000 yr and 5 kpc \citep[footnote 1 \& 2]{s2,g2,m1}
and for all of them, a SNR is detected \citep{g2}, 
and at least eight were observed historically \citep{s2}.  
While a missing historic observation is possible, a missing SNR is very unlikely.

Magnetar flares (soft gamma-ray repeaters or anomalous X-ray pulsars) were not yet considered: 
The largest flare observed was the X- and $\gamma$-ray flare of SGR 1806-20 on 2004 Dec 27 with peak 
energy ($3.7 \pm 0.9) \cdot 10^{46}$ erg/s at 15 kpc \citep{h2}, 
or E(event) = $2 \cdot 10^{46}$ erg at $8.7 \pm 1.7$ kpc \citep{b5}.  
If the AD 774/5 event were such a flare, it would have taken place at $\sim 5.5$ pc (Equ. 1 with $g = 1$), 
but there is no neutron star known within such a small distance \citep[footnote 2]{m1}
Even if a magnetar with $10^{16}$ G dipole field could produce an event with $10^{48}$ erg \citep{h2}, 
the distance of such a neutron star to produce the AD 774/5 event would have to be $\sim 39$ pc.  
A magnetar at that small distance would have been detected by the ROSAT all-sky X-ray survey: 
For a persistent bolometric luminosity (mostly X-rays) of $\sim (0.025-1.6) \cdot 10^{35}$ erg/s 
with typical observed spectral components of magnetars (blackbody with peak 
energy $k \cdot T = 0.4$ keV and power-law index $\sim 3$, footnote 2), 
we expect 150 to 800000 cts/s 
in the ROSAT energy band 0.1-2.4 keV at 10 to even 100 pc, i.e. easily detectable. 
Hence, we can exclude magnetar flares for the AD 774/5 event.

\section{A short gamma-ray burst}

Given that events on Earth as well as solar and stellar flares (M12) including neutron star
flares (see above) as well as unabsorbed (M12) and absorbed supernovae are very unlikely
to be the cause for the AD 774/5 cosmic-ray event (see above), we will now consider
a gamma-ray burst (GRB).
The observed duration and spectral hardness of GRBs 
allows to divide them into long ($\ge 2$s) and short ($\le 2$s) GRBs, the latter are harder regarding 
the spectrum (power-law with exponential cutoff) and are not related to SNe nor SNRs \citep{n1}. 
While long GRBs are caused by the core collapse of a very massive star,
short GRBs are explained by the merger of two compact objects \citep{n1}. 
A merger of two previously orbiting compact objects is the coalescence of a neutron star 
with either a black hole becoming a more massive black hole, or with another neutron star 
becoming either a relatively massive stable neutron star or otherwise a black hole, 
if the total mass exceeds the upper mass limit of neutron stars, 
somewhere between 2 and 3 M$_{\odot}$.
E.g., the merger of two magnetized neutron stars can produce a spinning black hole launching 
a relativistic jet as observed in short GRBs \citep{r2}, if the Earth is located in the jet. 
Let us now consider a short GRB.

\subsection{Energetics, time-scale, and spectrum}

A short GRB emits an isotropic equivalent energy of E(event) = $10^{49}$ to $10^{52}$ erg 
in the observed energy range 10 keV to 30 GeV \citep{n1,b3}, most or all in $\gamma$-rays ($g = 0.1$ to 1). 
We estimate the distance towards a short GRB from Equ. 1 to $d \simeq 0.1$ to $3.9$ kpc, i.e. within our Galaxy. 
Hence, the energetics of the $^{14}$C peak on Earth are consistent with a short GRB.  

Effects of nearby long GRBs on the Earth biosphere due to the direct hit ($5 \cdot 10^{51}$ erg/s for 10s)
on one half-sphere were found to be lethal within $\sim 2$ kpc \citep{m2,t2}.
This can be scaled to a short GRB with $10^{49}$ to $10^{52}$ erg.
Hence, for a short GRB within $\sim 1$ kpc, strong extinction effects are expected.
Because no extinction event was observed on Earth for AD 774/5,
the short GRB was more distant, probably $\sim 1$ to $4$ kpc.

A transient event is expected in the optical (macronova) from compact mergers with 
$M_{\rm V} = -15$ mag at peak \citep{m3,p2}.
This corresponds to $m_{\rm V} = -10$ (0.1 kpc) or $m_{\rm V} = -2$ mag (4 kpc) for negligible absorption. 
Hence, it may have been observable by naked eye, but only for up to one day, i.e. much shorter than a typical SN. 
If reports about such a sighting remain missing, it can be due to the short time-scale, strong absorption, 
bad weather, and/or sky location near the Sun and/or above unpopulated areas such as the Pacific during the 
short visibility period. A missing historical observation and a missing SNR are fully consistent with a short GRB. 

Since the peak of $^{14}$C observed in AD 774/5 is consistent with a sharp increase
within 0.1 to 1 yr (M12), a short GRB typically lasting less than 2s \citep{n1,r2}
and being undispersed in interstellar space is consistent with the observations regarding the short time-scale.

\begin{figure}
\includegraphics[width=0.35\textwidth,angle=270]{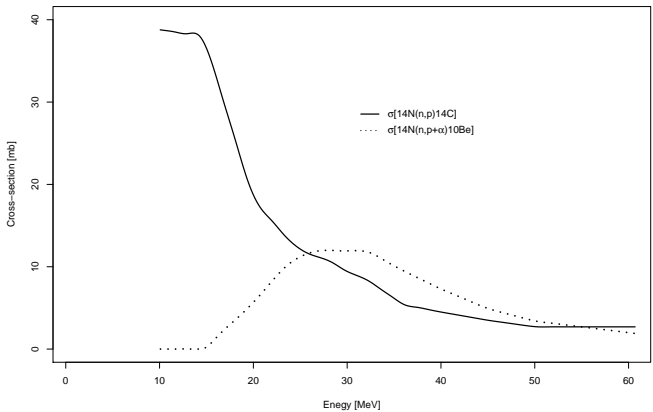}
\caption{Energy dependence of the relevant cross-sections for producing $^{14}$C and $^{10}$Be.
We plot the energy dependence of the cross-sections $\sigma$ of the reactions
$^{14}$N(n,p)$^{14}$C and $^{14}$N(n,p+$\alpha$)$^{10}$Be 
as full and dotted lines, respectively,
cross-section in milli-barn (mb, 1 barn is $10^{-28}$ m$^{2}$) versus energy in MeV.}
\label{fg1}
\end{figure}

\begin{figure}
\includegraphics[width=0.35\textwidth,angle=270]{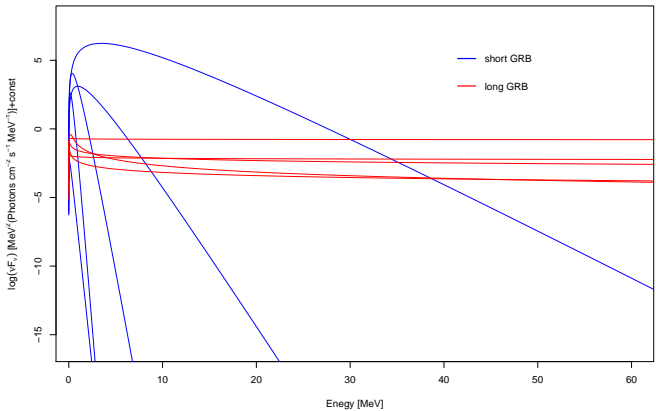}
\caption{Typical spectra of short and long gamma-ray bursts.
We plot the flux $\nu \cdot F_{\nu}$ (with frequency $\nu$) versus energy in MeV 
sampling the whole range of parameters observed,
i.e. several typical spectra as blue and red lines for short and long GRBs, respectively.
We plot the flux as log of $\nu \cdot F_{\nu}$ in MeV$^{2} \cdot$ photons $\cdot$ cm$^{-2} \cdot$ s$^{-1} \cdot$ MeV$^{-1}$
(plus arbitrary scaling due to normalization).
At the observable low $\gamma$-ray energies, we see that short GRBs 
are harder than long GRBs, hence the devision and naming.
We use Equ. 2 \& 3 and the Band function \citep{b2} to
estimate the production rate of $^{14}$C to $^{10}$Be from the input spectra for short and long GRBs.
Because long GRBs have a flat spectrum (Band function), they cannot reproduce
the observed production rate of $^{14}$C to $^{10}$Be ($\ge 270 \pm 140$), 
while short GRBs can reproduce the observed ratio, see Fig.~\ref{fg3}.}
\label{fg2}
\end{figure}

\begin{figure*}
\includegraphics[width=0.65\textwidth,angle=270]{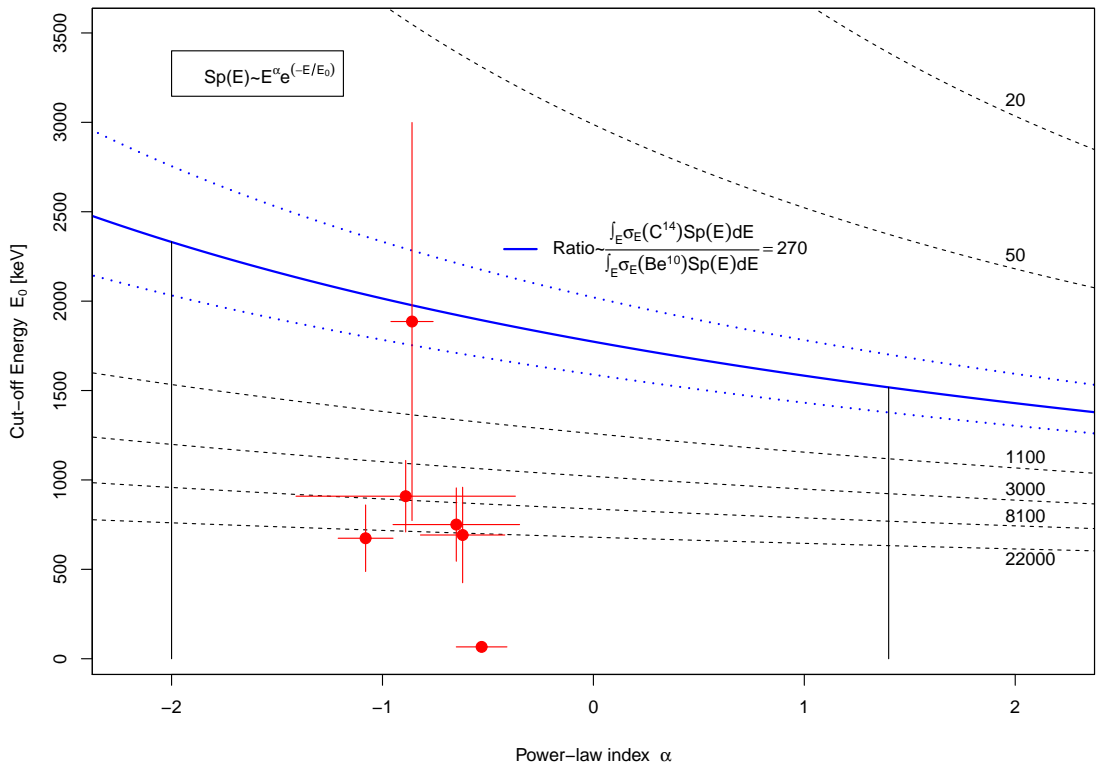}
\caption{The typical spectrum of short GRBs can explain the differential production
rates of $^{14}$C to $^{10}$Be observed to be at least $270 \pm 140$.
Given the typical energy spectrum of a short GRB with a power-law index
$\alpha = -2$ to $1.4$ (vertical black lines) and a cutoff energy
$E_{0} = 49$ to $1900$ keV \citep{n1,g1},
and also given the cross-sections
of the reactions $^{14}$N(n,p)$^{14}$C and $^{14}$N(n,p$\alpha$)$^{10}$Be from 10 to 60 MeV (footnote 3),
we estimated the expected fraction for a grid of $\alpha$ and $E_{0}$.
The observed lower limit on this fraction ($270 \pm 140$, $1~\sigma$ error)
is shown as blue curve with dotted lines as error margin.
The dashed black lines indicate the ratio of $^{14}$C to $^{10}$Be production
from low values in the upper right to high values in the lower left.
We also show the observed power-law indices and cutoff energies of
all six short GRBs with known redshifts \citep{g1}
as red dots with vertical and horizontal red $1~\sigma$ error bars.
Because the power-law index and the cutoff energy needed to
reproduce the observed lower limit of $270 \pm 140$ lies in the typical range of short GRBs
and because all data points of local short GRBs lie on or below the line for $270 \pm 140$,
the observed production ratio of $^{14}$C to $^{10}$Be in AD 774/5 is well consistent with a typical short GRB.}
\label{fg3}
\end{figure*}

Given the cross-sections\footnote{www-nds.iaea.org, \citep{d1,b7}}
of the relevant reactions producing $^{14}$C and $^{10}$Be (Fig.~\ref{fg1}) 
and the full range of observed spectral parameters of short \citep{n1,g1}
and long \citep[Band function,][]{b2}
GRBs (Fig.~\ref{fg2}), we computed the outcome (Fig.~\ref{fg3}).
We assume that the peaks in $^{14}$C ($19 \pm 4$ atoms/cm$^{2}$/s in $\le 1$ yr, M12),
and $^{10}$Be \citep[$30\%$ increase with 10 yr time resolution,][]{h1}
were both due to the same event, i.e. produced within one year.
Then, with the known background rates
for $^{14}$C (M12) and
for $^{10}$Be \citep{h1}, we conclude that $270 \pm 140$ (1$\sigma$ error) times more $^{14}$C was produced than $^{10}$Be,
above the respective backgrounds. Since some of the $^{10}$Be production may have been produced by
some other effects in that decade (lower time resolution), this ratio can be considered as a lower limit.
We can compare this to the expected outcome for a short GRB:
The typical energy spectrum of a short GRB is
\begin{equation}
Sp(E) \sim E^{\alpha} \times e^{(-E/E_{0})}
\end{equation}
with a power-law index $\alpha$ and a cutoff energy E$_{0}$  \citep{n1,g1}.
We obtained the cross-sections $\sigma$ for the reactions $^{14}$N(n,p)$^{14}$C and $^{14}$N(n,p+$\alpha$)$^{10}$Be
from E = 10 to 60 MeV (footnote 3). We show the dependence of the cross-sections
$\sigma$ on the energy E for both reactions in Fig.~\ref{fg1}. Then, we integrated the
cross-section $\sigma$ at that energy range E multiplied by the typical short GRB spectra
Sp(E) over the energy E from 10 to 60 MeV for both $^{14}$C and $^{10}$Be to obtain the expected ratio:
\begin{equation}
ratio = \frac{\int_{E} \sigma (^{14}C) \times Sp(E)~~dE}{\int_{E} \sigma (^{10}Be) \times Sp(E)~~dE}
\end{equation}
Given the definition of cross-section, both the nominator and the denominator in this equation
are proportional to the respective number of atoms produced.

We estimated the ratio as expected differential production rates for a grid of $\alpha$ and E$_{0}$
as observed for short GRBs, i.e. power law index $\alpha$ from $-2$ to 1.4
and cutoff energy E$_{0}$ from 49 to 1900 keV \citep{n1,g1},
using XSPEC. The results are shown 
in Fig.~\ref{fg3}.
We then also estimated the expected fraction of $^{14}$C to $^{10}$Be production over the backgrounds
for a long GRB with a Band function as spectrum \citep{b2}, i.e. a smoothly broken power-law.
Again, we used the full range of parameters observed for all those long GRBs, where all
three spectral parameters were obtained by good fits, namely for the 1st index $\alpha$
from $-1.40$ to 0.07, for the 2nd index $\beta$ from $-3.68$ to $-2.04$,
and for the cutoff energy E$_{0}$ from 52 to 2867 keV (not
redshifted) for long GRBs observed by FERMI or BATSE \citep{n1,g1,z1}.

Fig.~\ref{fg2} already shows qualitatively that a long GRB with a model of a smoothly broken power law (Band function)
cannot explain that the $^{14}$C production was so much larger than the $^{10}$Be production
given the energy-dependence of the cross-sections of those reactions.
Our detailed calculations show that a long GRB with the Band function would produce only up
to 18 times more $^{14}$C than $^{10}$Be. Only a short GRB can explain this observable
(ratio $\ge 270 \pm 140$).

\subsection{Rates of gamma-ray bursts and mergers}

The rate of $^{14}$C peaks (1 in 3000 yr) is not well constrained, but non-zero; 
the error of the rate is fully unconstrained. 
%
%

The rate of short GRBs is $8^{+5} _{-3}$ Gpc$^{-3}$ yr$^{-1}$ \citep{c1},
all observed and pointed at us.
Using then $\sim 0.003$ Mpc$^{-3}$ as number density of typical galaxies \citep{m2},
we have one short Galactic GRB in $375 ^{+225} _{-144}$ kyr beamed towards Earth.
If we restrict this estimate to the Galactic disk within 4 kpc, the observed rate is 10 times
lower, i.e. one nearby Galactic GRB beamed at us in $3750 ^{+2250} _{-1442}$ kyr.
Given the sensitivity limits of detectors like BATSE and SWIFT, this is a highly uncertain lower limit.
While this is not consistent with 1 event in 3000 yr within $1~\sigma$, it is consistent within $2.6~\sigma$.
Given that the error of the $^{14}$C event rate is unknown, it is consistent with the rate of 
local short GRBs even within $\le 2.6~\sigma$.

Even though the possible connection between short GRBs and compact mergers are not of central
importance for our arguments, because we use only real observables of short GRBs, we will now
consider the rates of mergers of compact objects.
From the three known double neutron stars, one can expect the rate of mergers per galaxy to be
3 to 190 Myr$^{-1}$ ($1~\sigma$ error range) with the mean being 13 Myr$^{-1}$ \citep{k1}.
From the initial mass function and, hence, birth rate of massive stars that can become neutron
stars, then taking into account the multiplicity rate, evolution, and interaction of massive stars,
one can predict 0.3 to 50 mergers per galaxy Myr$^{-1}$ ($1~\sigma$ error range) with the mean being 15 Myr$^{-1}$ \citep{d3}.
Thus, at most we expect 190 mergers per galaxy Myr$^{-1}$ or up to one merger in $\sim 5263$ yr.
If such a merger would be observable as short GRB, one would have to correct the rate for the
beaming fraction $f=0.01$ to 0.13 for short GRBs \citep{r2}.
For $f=0.13$, one would then expect up to one merger in $\sim 40000$ yr
(within $1~\sigma$ error bars), pointed towards Earth as short GRB
(ten times less within 4 kpc).

Because of sensitivity limits of observational techniques, both the observed multiplicity
rate of massive stars and the estimated number of double neutron stars and mergers are lower limits.
We can add the rates of mergers between neutron stars and black holes
and between two black holes.

Neither the highly uncertain rates of observed short GRBs nor of neutron star mergers 
are consistent with the observed rate of the $^{14}$C event 
(one event in 3000 yr) within $1~\sigma$, but all are consistent within $\le 2.6~\sigma$.
Furthermore, a short GRB is the only known phenomenon that can provide correct energetics,
correct spectrum, and correct time-scale for the observed event; it also does not produce a typical SN
light curve for several month nor a detectable SNR nor a mass extinction event on Earth, which are all missing.
If the AD 774/5 event was a short GRB and if the probability to observe one
Galactic GRB within 3000 yr is too small, one would have to conclude that there are more
(fainter) short GRBs than observed so far, and/or that there is
another astrophysical population contributing to short GRBs, which was not yet fully recognized.

Short GRBs with extended emission may partly be due to either an accretion-induced
collapse of a white dwarf or the merger of two white dwarfs \citep{b4,b6,m4}.
In such an event, a magnetar can form \citep{b6}.
The rate of mergers of two white dwarfs with a total mass above the Chandrasekhar mass limit of
$\sim 1.4$ M$_{\odot}$ has been estimated to be 
$1.0 ^{+1.6} _{-0.6} \times 10^{-14}$ 
per M$_{\odot}$
with $1 \sigma$ error bars \citep{b1}.
This corresponds to only about one tenth of the SN Ia rate, so that the merger of two white dwarfs
with super-Chandrasekhar mass cannot explain all SN Ia events. However, it is suggested that such super-Chandrasekhar
mergers can be observed as short GRBs \citep{b4,b6,m4}.
For our Galaxy with $\sim 10^{11}$ M$_{\odot}$,
we obtain a rate of $3.0 ^{+4.8} _{-1.8}$ mergers of white dwarf binaries with super-Chandrasekhar mass in 3000 yr.
If we assume that such an event can be observed as short GRB with extended emission with a beaming factor
of up to $f = 0.25$ \citep{b6}, and if we also restrict the rate to the disk within 4 kpc
(the maximum distance of a short GRB to explain the AD 774/5 event),
we expect $0.08 ^{+0.12} _{-0.04}$ such mergers in 3000 yr.
Since the error of the rate of $^{14}$C events (as in AD 774/5) is unconstrained,
one cannot claim that the rates of white dwarf binary mergers with super-Chandrasekhar mass (pointing towards
us a short GRB) and the $^{14}$C event rate are inconsistent.

Since short GRBs with extended emission may have lower total energies \citep{b6}, 
which can be below the BATSE and SWIFT sensitivity limits, they may often remain undetected,
so that their observable rates could be underestimated.
There is evidence for short GRBs with lower energies and 
their rate is probably much higher \citep{l1,t1,n2}.
Given the discussion of the rates, we can speculate 
that some short GRBs, like possibly one in AD 774/5, are due to an accretion-induced
collapse of a white dwarf or the merger of two white dwarfs, and that such (possibly frequent) short 
GRBs typically have energies below the current sensitivity limit of $10^{49}$ erg and possibly wide
beaming angles; they may produce a neutron star, but no SNR.

\section{Concluding remarks}

A long GRB can be accompanied by a SN and a SNR, but none were observed for AD 774/5; 
this could be due to strong absorption (no optical sighting of the SN) and large distance 
(faint SNR with very small angular extension on sky). 
With the typical isotropic equivalent energy output E(event) = $10^{52}$ to $10^{54}$ erg of a long GRB 
in $\gamma$-rays of 10 keV to 30 GeV and $g = 0.1$ to 1 \citep{n1}, we estimate its putative distance towards Earth 
from Equ. 1 to $d \simeq 1$ to $39$ kpc, i.e. in our Galaxy 
or the neighbouring Canis Major, Sagittarius, or Ursa Major II dwarf galaxies. 
To avoid a historical sighting of a SN brighter than $m \simeq 2$ mag \citep{s2} at the minimum distance of 1 kpc, 
one would need an absorption of $A_{\rm V} = 12.5$ mag for a peak absolute magnitude of $M \simeq -21$ mag for 
a collapsar/hypernova \citep{r3}. An area of $66$ deg$^{2}$ 
has an absorption of $A_{\rm V} \ge 12.5$ mag \citep{d2}, an even smaller area for clouds within 1 kpc. 
There is no such SNR detected behind these areas \citep[footnote 1]{g2}.
Considering also that long GRBs are 20 times less frequent than short GRBs \citep{n1},
a long GRB behind such strong absorption is then $\ge 12727$ times less likely 
than a short GRB to explain the $^{14}$C peak in AD 774/5. 
Also, sampling the whole observed range of spectral parameters of long GRBs 
with their smoothly broken power law or Band function \citep{n1,g1,b2,z1},
we cannot explain the differential $^{14}$C to $^{10}$Be production observed 
(Fig.~\ref{fg1}). Hence, a short GRB remains the only plausible explanation for the $^{14}$C peak in AD 774/5.

We list in Table 1 all known neutron stars with characteristic age $\le 25000$ yr
\citep[footnote 2]{m1}
at distances
from 1 to 4 kpc, but without any known SNR (Green 2009; footnote 1).
The list includes two Anomalous X-ray Pulsars (AXPs), one Fermi Gamma-ray Large area telescope pulsar (FGL),
and one Soft Gamma-ray Repeater (SGR).
We use a larger pulsar age upper limit (25000 yr) than the time since AD 774/5,
because characteristic ages 
are usually upper limits and can be 20 times larger than the true age \citep{k2}.
We include SGR 0418+5729 with an upper limit for the period derivative and, hence,
a lower limit on the age (which is below 25000 yr).
If a neutron star was formed in AD 774/5, it is also possible that it was not yet discovered,
e.g. because of misdirected pulsar beaming, or that distance and/or age have not yet been determined.
For the five pulsars listed here, one should obtain deep X-ray, $\gamma$-ray, H$\alpha$, and radio observations
to search for SNRs: If a SNR can be excluded in one of them, that pulsar may be a good candidate
for the product of the AD 774/5 event. Three of the five neutron stars listed are
AXPs or SGRs, which can form by a short GRB with extended emission \citep{b4,b6,m4}

\begin{table}
{\bf Table 1: Young neutron stars at 1-4 kpc without SNR.} We list pulsar name with position for J2000.0, rotation/pulse period $P$, period derivative $\dot P$,
distance, and characteristic age $\tau = 1/2 \cdot (P/\dot P)$, data from \citet{m1} or McGill SGR/AXP catalog (footnote 2). \\
\begin{tabular}{lrrccl} \hline 
Name           & Period  & P-dot        & Dist.    & Age      & Re- \\
J2000.0        & P [s]   & [s/s]        & [kpc]    & [kyr]    & mark     \\ \hline
SGR 0418+5729  & 9.0784  & $\le$ 0.0006 & $\sim$ 2 & $\ge$ 24 & \\
PSR J1048-5832 & 0.1237  & 9.6e-14      & 2.98     & 20.3     & FGL \\
PSR J1708-4009 & 11.0013 & 1.9e-11      & 3.08     & 9.01     & AXP \\
PSR J1740-3015 & 0.6069  & 4.6e-13      & 3.28     & 20.6     &     \\
PSR J1809-1943 & 5.5404  & 7.8e-12      & 3.57     & 11.3     & AXP \\ \hline
\end{tabular}
\end{table}

In summary, all observables of the $^{14}$C peak in AD 774/5 are consistent
with a Galactic short GRB at 1-4 kpc: Sufficient energetics, correct spectrum, and correct time-scale, 
also neither a SN nor a SNR nor a mass extinction event.
The only assumptions made were the following: From comparing their $^{14}$C tree ring data with a 
model of incorporation of $^{14}$C into the biosphere, M12 concluded that the event duration was
one year or shorter, and they could then derive the $^{14}$C flux and energy deposited on Earth
which was also used in our work;
we also used the $^{10}$Be flux observed in the same decade \citep{h1} and assumed
that it was produced by the same event \citep[see e.g.][]{s3}
to derive the differential $^{14}$C to $^{10}$Be production rate;
we derived a lower limit to the differential production rate, because some of the $^{10}$Be
observed in that decade could have been produced by other effects.
The derived lower limit was then found to be inconsistent
with long GRBs, but fully consistent with the spectra of short GRBs.
Rates of short GRBs and neutron star mergers are marginally consistent with one event in 3000 yr,
but the error of the rate of the AD 774/5 event is unknown.
The merger of two white dwarfs with super-Chandrasekhar mass
or an accretion-induced collapse of a white dwarf
producing a short GRB
(with $\le 10^{49}$ erg at $\le 1$ kpc) should also be considered. The product could be a 
neutron star without SNR, so that our conclusions are testable.

\section*{Acknowledgments}
We would like to thank the German national science foundation DFG
(Deutsche Forschungsgemeinschaft) for financial support through the collaborative research center
Sonderforschungsbereich SFB TR 7 Gravitational Wave Astronomy sub-project C7.
We used the online catalog of Supernova Remnants by D. Green,
the Chandra supernova remnant catalog maintained by F. Seward,
the ATNF online catalog of pulsars maintained by G.B. Hobbs and R.N. Manchester,
and the McGill online catalog of SGRs and AXPs maintained by the McGill Pulsar Group.
We also thank Kazuhito Dobashi for information about the size of the sky area, where the
extinction is $A_{\rm V} \ge 13$ or $\ge 12.5$ mag, considering the whole sky.

\end{document}